\documentclass[aps,pra,twocolumn,superscriptaddress]{revtex4-1}%
\usepackage{graphics}
\usepackage{amsmath}
\usepackage{graphicx}
\usepackage{amsfonts}
\usepackage{amssymb}
\usepackage{float}
\usepackage{longtable}
\usepackage{epsfig}
\usepackage{latexsym}
\usepackage{theorem}
\usepackage{tikz}
\usepackage{bbm}
\usepackage{bm}
\usepackage{braket}
\usepackage{psfrag}

\newtheorem{theorem}{Theorem}
\newtheorem{lemma}[theorem]{Lemma}

\def\^PT{^{\mathrm{PT}}}
\begin{document}
\title{Converse bounds for quantum and private\\communication over Holevo-Werner channels}
\author{Thomas P. W. Cope}
\affiliation{Computer Science and York Centre for Quantum Technologies, University of York,
York YO10 5GH, UK}
\author{Kenneth Goodenough}
\affiliation{QuTech, Delft University of Technology, Lorentzweg 1, 2628 CJ Delft, The Netherlands}
\author{Stefano Pirandola}
\affiliation{Computer Science and York Centre for Quantum Technologies, University of York,
York YO10 5GH, UK}

\begin{abstract}
Werner states have a host of interesting properties, which often serve to
illuminate the unusual properties of quantum information. Starting from these
states, one may define a family of quantum channels, known as the
Holevo-Werner channels, which themselves afford several unusual properties. In
this paper we use the teleportation covariance of these channels to upper
bound their two-way assisted quantum and secret-key capacities. This bound may
be expressed in terms of relative entropy distances, such as the relative
entropy of entanglement, and also in terms of the squashed entanglement. Most
interestingly, we show that the relative entropy bounds are strictly
subadditive for a sub-class of the Holevo-Werner channels, so that their
regularisation provides a tighter performance. These information-theoretic
results are first found for point-to-point communication and then extended to
repeater chains and quantum networks, under different types of routing strategies.

\end{abstract}
\maketitle

\section{Introduction}

The area of quantum information and
computation~\cite{NielsenChuang,Hayashi,HolevoBOOK,RMP,HayashiINTRO}
is one of the fastest growing fields. Understanding how quantum
information is transmitted is necessary not only for the
development of a future quantum
Internet~\cite{Kimble2008,HybridINTERNET,ref1,ref3,ref4,ref5,ref6,Meter}
but also for the construction of practical quantum key
distribution (QKD)~\cite{crypt1,crypt2,crypt3,crypt4} networks.
Motived by this, there is much interest in trying to establish the
optimal performance in the transmission of quantum bits (qubits),
entanglement bits (ebits) and secret bits between two remote
users. This is a theoretical framework which is a direct quantum
generalization of Shannon's theory of
information~\cite{Shannon,Cover&Thomas}. In the quantum setting,
there are different types of maximum rates, i.e., capacities, that
may be defined for a given quantum channel. These include the
classical capacity (transmission of classical bits), the
entanglement distribution capacity (distribution of ebits), the
quantum capacity (transmission of qubits), the private capacity
(transmission of private bits), and the secret-key capacity
(distribution of secret bits). All these capacities may be defined
allowing side local operations (LOs) and classical communication
(CC) either one-way or two-way between the remote parties.

We shall focus on the use of LOs assisted by two-way CC, also known as
\textquotedblleft adaptive LOCCs\textquotedblright. The maximization over
these types of LOCCs leads to the definition of corresponding two-way assisted
capacities. In particular, in this work we are interested in the two-way
quantum capacity $Q_{2}$ (which is equal to the two-way entanglement
distribution capacity $D_{2}$) and the secret-key capacity $K$ (which is equal
to the two-way private capacity $P_{2}$). Generally, these capacities are
extremely difficult to calculate because they involve quantum protocols based
on adaptive LOCCs, where input states and the output measurements are
optimized in an interactive way by the two remote parties. Similar adaptive
protocols may be considered in other tasks, such as quantum hypothesis
testing~\cite{Harrow,PirCo,PBT} and quantum
metrology~\cite{PirCo,Rafal,ReviewMETRO,adaptive2,ada3,ada4,ada5,HWmetro}.

Building on a number of preliminary
tools~\cite{B2main,Gatearray,SamBrassard,HoroTEL,Gottesman,SougatoBowen,Knill,WernerTELE,Cirac,Aliferis,Niset,MHthesis,Wolfnotes,Leung}
and generalizing ideas therein to arbitrary dimension and arbitrary tasks,
Ref.~\cite{Stretching} showed how to use the LOCC simulation~\cite{NoteLOCC}
of a quantum channel to reduce an arbitrary adaptive protocol into a simpler
block version. More precisely, Ref.~\cite{Stretching} showed how the suitable
combination of an adaptive-to-block reduction (teleportation stretching) with
an entanglement measure, such as the relative entropy of entanglement
(REE)~\cite{RMPrelent,VedFORMm,Pleniom}, allows one to reduce the expression
of $Q_{2}$ and $K$ to a computable single-letter version. In this way,
Ref.~\cite{Stretching} established the two-way capacities of several quantum
channels, including the bosonic lossy channel~\cite{RMP}, the quantum-limited
amplifier, the dephasing and the erasure channel~\cite{NielsenChuang}. The
secret-key capacity of the erasure channel was also established in a
simultaneous work~\cite{GEW} by using a different approach based on the
squashed entanglement~\cite{squash}, which also appears to be powerful in the
case of the amplitude damping channel~\cite{GEW,Stretching}. Note that, prior
to these results, only the $Q_{2}$ of the erasure channel was
known~\cite{ErasureChannelm}.

One of the golden rules to apply the previous techniques is teleportation
covariance, first considered for discrete variable (DV)
channels~\cite{MHthesis,Wolfnotes,Leung} and then extended to any dimension,
finite or infinite~\cite{Stretching}. This is the property of a quantum
channel to \textquotedblleft commute\textquotedblright\ with the random
unitaries of quantum teleportation~\cite{tele,teleCV,telereview,teleCV2}.
Because the Holevo-Werner (HW) channels~\cite{counter,Fanneschannel} are
teleportation covariant, we may therefore apply the previous reduction tools
and bound their two-way assisted capacities, $Q_{2}$ and $K$, via
single-letter quantities. These channels are particularly interesting because
the resulting upper bounds, based on relative entropy distances (such as the
REE), are generally non-additive. In fact, we show a regime of parameters
where a multi-letter bound is strictly tighter than a single-letter one.

As a result of this sub-additivity, the regularisation of the
upper bound needs to be considered for the capacities $Q_{2}$ and
$K$ of these channels. This is a property that the HW channels
inherit from their Choi matrices, the Werner states~\cite{Werner}.
Recall that these states may be entangled, yet admit a local model
for all measurements~\cite{Werner,Barrett}. They disproved the
additivity of REE~\cite{VolWer} (which is the main property
exploited here), and they are also conjectured to prove the
existence of negative partial transpose undistillable
states~\cite{NPTstates}.

Another interesting finding is that bounds which are based on the squashed
entanglement compete with the REE bounds in a way that there is not a clear
preference among them. In fact, we find that the secret-key capacity of an
HW\ channel is better bounded by the REE or the squashed entanglement
depending on the value of its main defining parameter. This is a feature which
has never been observed for another quantum channel so far.

The structure of this paper is as follows. We begin in Sec.~\ref{Wer_preli} by
introducing the mathematical description of both Werner states and HW
channels. In Sec.~\ref{REEsec} we review the notions of relative entropy
distance with respect to separable states and partial positive transpose (PPT)
states, also discussing their regularised versions. In Sec.~\ref{sec:TC}, we
compute the REE for the overall state consisting of two identical Werner
states, discussing the strict subadditivity of the REE for a subclass of the
family. Then, in Sec.~\ref{Capacities} we give our upper bounds to the $Q_{2}$
and $K$ of the HW channels, which too exhibits the subadditivity property.
Here we also prove a general upper bound for the $Q_{2}$ of any teleportation
covariant channel (at any dimension). In Sec.~\ref{SECnet}\ we extend the
results to repeater chains and quantum networks connected by HW channels. We
then conclude and summarize in Sec.~\ref{Werconclu}.

\begin{figure*}[th]
\begin{center}%
\begin{tabular}
[c]{c|c|c|c|c|c}%
~Representation~ & ~Variable~ & State & $~%
\begin{array}
[c]{c}%
\text{Separable}\\
\text{Extreme}%
\end{array}
~$ & ~Boundary~ & $~%
\begin{array}
[c]{c}%
\text{Entangled}\\
\text{Extreme}%
\end{array}
~$\\\hline
$\alpha$-rep & $\alpha$ & $\frac{1}{d^{2}-d\alpha}\left(  \mathbb{I}%
-\alpha\mathbb{F}\right)  $ & $-1$ & $\frac{1}{d}$ & $1$\\
Weighting rep & $p$ & $\frac{1-p}{d^{2}+d}\left(  \mathbb{I}+\mathbb{F}%
\right)  +\frac{p}{d^{2}-d}\left(  \mathbb{I}-\mathbb{F}\right)  $ & $0$ &
$\frac{1}{2}$ & $1$\\
Expectation rep & $\langle\mathbb{F}\rangle=\eta$ & $~\frac{1}{d^{3}-d}\left[
(d-\eta)\mathbb{I}+(d\eta-1)\mathbb{F}\right]  ~$ & $1$ & $0$ & $-1$\\
Anti-rep & $t$ & $t\frac{\mathbb{I}-d\mathbb{F}}{d^{2}(d-1)}+\frac{\mathbb{I}%
}{d^{2}}$ & $-\frac{1}{d-1}$ & $\frac{1}{d+1}$ & $1$%
\end{tabular}
\end{center}
\caption{The various ways in which the set of Werner states of dimension $d$
may be parametrised. All of these are equivalent and may be transformed
between. Here $\mathbb{I}$ is the $d^{2}$-dimensional identity operator and
$\mathbb{F}$ is the flip operator.}%
\label{reps}%
\end{figure*}

\section{Werner states and Holevo-Werner channels\label{Wer_preli}}

Werner states are an important family of quantum states which are generally
defined over two qudits of equal dimension $d$. They have the peculiar
property to be invariant under unitaries $U_{d}$ applied identically to both
subsystems, i.e., they satisfy the fixed-point equation%
\begin{equation}
(U_{d}\otimes U_{d})\rho(U_{d}^{\dagger}\otimes U_{d}^{\dagger})=\rho.
\label{invar}%
\end{equation}
There exists several parametrisations of this family as also shown in
Fig.~\ref{reps}. We shall use the \textquotedblleft expectation
representation\textquotedblright, where the Werner state $W_{\eta,d}$ is
parametrised by $\eta\in\lbrack0,1]$ which is defined by the mean value
\begin{equation}
\eta:=\mathrm{Tr}[W_{\eta,d}\mathbb{F}], \label{Tracecon}%
\end{equation}
where $\mathbb{F}$ is the flip operator acting on two qudits in the
computational basis $\left\{  \ket{i}\right\}  _{i=0}^{d-1}$, i.e.,
\begin{equation}
\mathbb{F}:=\sum_{i,j=0}^{d-1}\ket{ij}\bra{ji}.
\end{equation}
If $\eta$ is negative (non-negative), then the Werner state is entangled
(separable). One also has an explicit formula for $W_{\eta,d}$ as a linear
combination of the $\mathbb{F}$ operator and the $d^{2}$-dimensional identity
operator $\mathbb{I}$, i.e.,
\begin{equation}
W_{\eta,d}=\frac{(d-\eta)\mathbb{I}+(d\eta-1)\mathbb{F}}{d^{3}-d}.
\label{stateform}%
\end{equation}

As already mentioned before, Werner states are of much interest to quantum
information theorists due to their properties. For $d\geq3$ there are Werner
states which are entangled, yet admit a local model for all
measurements~\cite{Werner,Barrett}. In particular, the extremal entangled
Werner state $W_{-1,d}$ was used to disprove the additivity of the
REE~\cite{VolWer}. A useful property of the Werner states is that, for a given
dimension, they are simultaneously diagonalisable, i.e., they share a common
eigenbasis. A Werner state $W_{\eta,d}$ has $n_{+}$ ($n_{-}$) eigenvectors
with eigenvalue $\gamma_{+}$ ($\gamma_{-}$), where $n_{\pm}:=d(d\pm1)/2$ and
$\gamma_{\pm}:=(1\pm\eta)[d(d\pm1)]^{-1}$.

\begin{figure}[ptb]
\begin{center}
\includegraphics[width=0.23\textwidth]{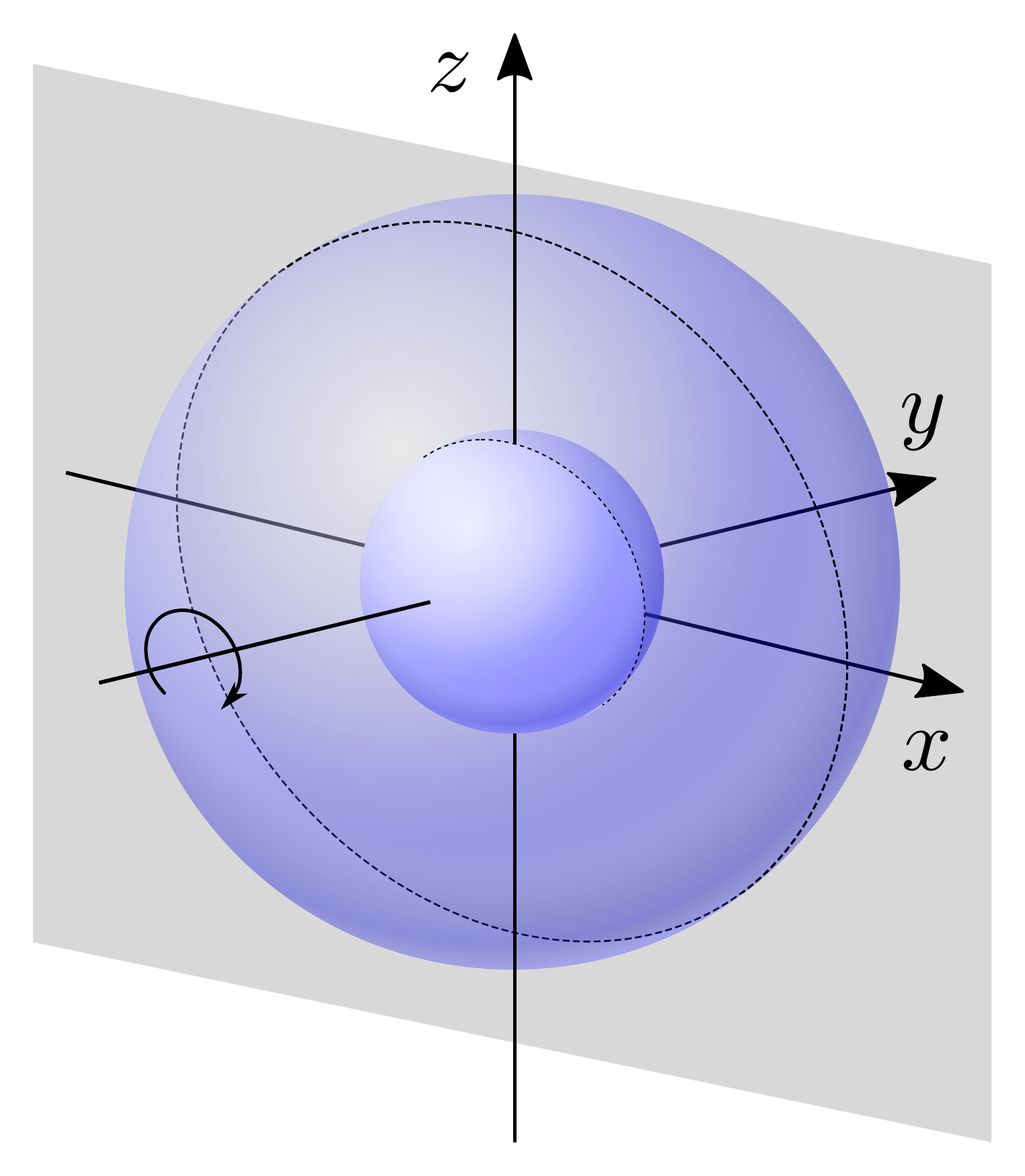}
\end{center}
\par
\vspace{-0.2cm}\caption{An illustration of the qubit HW channel ($d=2$). The
Bloch sphere is shrunk by a factor of $|\frac{2\eta-1}{3}|$, with the state
reflected in the $x$-$z$ axis for $\frac{2\eta-1}{3}>0$, and rotated by $\pi$
around the $y$ axis for $\frac{2\eta-1}{3}<0$.}%
\label{HW2}%
\end{figure}

Closely linked with Werner states are the HW
channels~\cite{counter,Fanneschannel}.\ These are defined as those channels
$\mathcal{W}_{\eta,d}$ whose Choi matrices are Werner states $W_{\eta,d}$. In
other words, we have
\begin{equation}
W_{\eta,d}:=\mathbf{I}\otimes\mathcal{W}_{\eta,d}\left(
\ket{\Phi}\bra{\Phi}\right)  ,
\end{equation}
where $\mathbf{I}$\ is the $d$-dimensional identity map and
$\ket{\Phi}=d^{-1/2}\sum_{i=0}^{d-1}\ket{ii}$ is a maximally-entangled state.
This is a family of quantum channels whose action can be expressed as%
\begin{equation}
\mathcal{W}_{\eta,d}\left(  \rho\right)  :=\frac{(d-\eta)\mathbf{I}%
+(d\eta-1)\rho^{T}}{d^{2}-1}, \label{channeldef}%
\end{equation}
where $T$ is transposition (see Fig.~\ref{HW2} for a representation in the
specific case $d=2$). It is known that the minimal output entropy of the HW
channels is additive~\cite{Fanneschannel}, and the extremal HW channel (for
$\eta=-1$) is a counterexample of the additivity of the minimal Reny\'{\i}
entropy~\cite{counter}. HW channels were also studied by Ref.~\cite{Leung}\ in
relation to forward-assisted quantum error correcting codes and
superactivation of quantum capacity.

An important property of the HW channels is their \emph{teleportation
covariance}. A quantum channel $\mathcal{E}$ is called \textquotedblleft
teleportation covariant\textquotedblright\ if, for any teleportation unitary
$U$, there exists some unitary $V$ such that~\cite{Stretching}
\begin{equation}
\mathcal{E}\left(  U\rho U^{\dagger}\right)  =V\mathcal{E}\left(  \rho\right)
V^{\dagger},
\end{equation}
for any state $\rho$. The teleportation unitaries referred to here are the
Weyl-Heisenberg generalisation of the Pauli matrices~\cite{NielsenChuang}.
Note that the output unitary $V$ may belong to a different representation of
the input group. For an HW channel $\mathcal{W}_{\eta,d}$, it is easy to see
that we may write
\begin{equation}
\mathcal{W}_{\eta,d}(U_{d}\rho U_{d}^{\dagger})=U_{d}^{\ast}\mathcal{W}%
_{\eta,d}(\rho)(U_{d}^{\ast})^{\dagger},
\end{equation}
for an arbitary unitary $U_{d}$. This comes from Eq.~(\ref{channeldef}) and
noting that $\mathbf{I}=U_{d}^{\ast}\mathbf{I}(U_{d}^{\ast})^{\dagger}$ and
$(U_{d}\rho U_{d}^{\dagger})^{T}=U_{d}^{\ast}\rho^{T}(U_{d}^{\ast})^{\dagger}$.

\section{Relative entropy distances\label{REEsec}}

An important functional of two quantum states $\rho$ and $\sigma$ is their
relative entropy, which is defined as
\begin{equation}
S(\rho||\sigma)=\mathrm{Tr}\left(  \rho\mathrm{log}_{2}\rho-\rho
\mathrm{log}_{2}\sigma\right)  .
\end{equation}
This is the basis for defining relative entropy distances. Given any compact
and convex set of states $S$ (containing the maximally mixed state), the
relative entropy distance of a state $\rho$ from this set is defined
as~\cite{Horos}%
\begin{equation}
E_{S}\left(  \rho\right)  :=\inf_{\sigma\in S}S(\rho||\sigma).
\end{equation}
This is known to be asymptotically continuous~\cite{Horos,Donald}. One
possible choice for $S$ is the set of separable (\textrm{SEP}) states, in
which case we have the REE~\cite{RMPrelent,VedFORMm,Pleniom}%
\begin{equation}
E_{R}\left(  \rho\right)  :=\inf_{\sigma\in\mathrm{Sep}}S(\rho||\sigma).
\end{equation}

Another possible choice is the set of PPT states, in which case we have the
relative entropy distance with respect to PPT states, which we denote by RPPT.
This is defined as%
\begin{equation}
E_{P}\left(  \rho\right)  :=\inf_{\sigma\in\mathrm{PPT}}S(\rho||\sigma),
\end{equation}
which coincides with the Rains' bound~\cite{Rains,Rains2} when $\rho$ is a
Werner state, as shown in Ref.~\cite{Audenert}. Recall that a PPT state
$\sigma$ is such that $\sigma^{\mathrm{PT}}$ has non-negative eigenvalues
(where $\mathrm{PT}$ is transposition over the second subsystem only). This is
a \emph{necessary} condition for $\sigma$ to be separable, but is \emph{not
sufficient}, unless $\sigma$ is a 2-qubit or qubit-qutrit state. Thus, in
general, we have
\begin{equation}
E_{P}(\rho)\leq E_{R}(\rho).
\end{equation}

Both the measures here defined are subadditive, i.e., they have the following
property under tensor product,
\begin{equation}
E_{R(P)}^{2}(\rho):=\frac{E_{R(P)}\left(  \rho^{\otimes2}\right)  }{2}\leq
E_{R(P)}\left(  \rho\right)  .
\end{equation}
It was shown that there exist states which are \emph{strictly} subadditive
($<$). In fact, for $d>2$, Ref.~\cite{VolWer} proved that
\begin{equation}
E_{R(P)}^{2}\left(  W_{-1,d}\right)  <E_{R(P)}(W_{-1,d}).
\end{equation}
This motivates the definition of the regularised quantities
\begin{equation}
E_{R(P)}^{\infty}\left(  \rho\right)  =\lim_{n\rightarrow\infty}\frac
{E_{R(P)}\left(  \rho^{\otimes n}\right)  }{n}\leq E_{R(P)}\left(
\rho\right)  ,
\end{equation}
i.e., the regularised REE $E_{R}^{\infty}$ and RPPT $E_{P}^{\infty}\leq
E_{R}^{\infty}$.

For an entangled Werner state, the closest separable and PPT state (for one
copy) is the boundary Werner separable state $W_{0,d}$, so that~\cite{VolWer}%
\begin{align}
&  E_{R(P)}\left(  W_{\eta,d}\right)  \label{onecopyREE}\\
&  =%
\begin{cases}
0 & \text{ if }\eta\geq0,\\
\frac{1+\eta}{2}\mathrm{log}_{2}\left(  1+\eta\right)  +\frac{1-\eta}%
{2}\mathrm{log}_{2}\left(  1-\eta\right)   & \text{ if }\eta\leq0.
\end{cases}
\nonumber
\end{align}
Note that the one-copy quantity $E_{R(P)}\left(  W_{\eta,d}\right)  $ does not
depend on the dimension $d$. Then, for Werner states, the regularised RPPT
$E_{P}^{\infty}$ is known~\cite{Audenert} and reads
\begin{align}
&  E_{P}^{\infty}\left(  W_{\eta,d}\right)  \label{PPTres}\\
= &
\begin{cases}
0 & \text{ if }\eta\geq0,\\
\frac{1+\eta}{2}\mathrm{log}_{2}\left(  1+\eta\right)  +\frac{1-\eta}%
{2}\mathrm{log}_{2}\left(  1-\eta\right)   & \text{ if }-\frac{2}{d}\leq
\eta\leq0,\\
\mathrm{log}_{2}\left(  \frac{d+2}{d}\right)  +\frac{1+\eta}{2}\mathrm{log}%
_{2}\left(  \frac{d-2}{d+2}\right)   & \text{ if }\eta\leq-\frac{2}{d}.
\end{cases}
\nonumber
\end{align}
From the previous equation, we see that we have strict subadditivity
$E_{P}^{\infty}\left(  W_{\eta,d}\right)  <E_{P}\left(  W_{\eta,d}\right)  $
in the region $\eta<-2/d$. Note that, in the region $-2/d\leq\eta\leq0$, the
REE is additive, and that the REE, the RPPT, and their regularised versions
all coincide. In fact, using the previous results, one has%
\begin{align}
E_{R}\left(  W_{\eta,d}\right)   &  =E_{P}\left(  W_{\eta,d}\right)
=E_{P}^{\infty}\left(  W_{\eta,d}\right)  \nonumber\\
&  \leq E_{R}^{\infty}\left(  W_{\eta,d}\right)  \leq E_{R}\left(  W_{\eta
,d}\right)  .
\end{align}

\section{Relative entopy distance of a two-copy Werner state\label{sec:TC}}

One of the results of Ref.~\cite{VolWer} was to show that the
closest state $\sigma$ minimizing $E_{R(P)}(W_{\eta,d}^{\otimes
n})$ is invariant under the following transformation
\begin{equation}
U_{d}^{1}\otimes U_{d}^{1}\otimes\ldots U_{d}^{n}\otimes U_{d}^{n}\left(
\sigma\right)  (U_{d}^{1}\otimes U_{d}^{1}\otimes\ldots U_{d}^{n}\otimes
U_{d}^{n})^{\dagger},
\end{equation}
where each $U_{d}^{i}\otimes U_{d}^{i}$ acts on the $d\times d$ Hilbert space
occupied by the $i^{\mathrm{th}}$ copy of $W_{\eta,d}$. States which are
invariant under this action are of the form
\begin{align}
\sigma_{\mathbf{x}}^{n}  &  =x_{0}W_{-1,d}^{\otimes n}\nonumber\\
&  +\frac{x_{1}}{n}\left(  W_{-1,d}^{\otimes n-1}\otimes W_{1,d}+\ldots
W_{1,d}\otimes W_{-1,d}^{\otimes n-1}\right) \nonumber\\
&  +\ldots+\frac{x_{k}}{\binom{n}{k}}\left(  W_{-1,d}^{n-k}\otimes W_{1,d}%
^{k}\ldots W_{1,d}^{k}\otimes W_{-1,d}^{n-k}\right) \nonumber\\
&  +\ldots+x_{n}W_{1,d}^{\otimes n},
\end{align}
where $\mathbf{x}=\left(  x_{0},x_{1},\ldots,x_{n}\right)  ^{T}$ is a vector
of probabilities, i.e., $x_{i}\geq0$ and $\sum_{i=0}^{n}x_{i}=1$. We also have
an explicit condition of $\mathbf{x}$ to ensure that $\sigma_{\mathbf{x}}^{n}$
is PPT. This is~\cite{Audenert}
\begin{equation}
\left(
\begin{array}
[c]{cc}%
-1 & 1\\
1 & \frac{d-1}{d+1}%
\end{array}
\right)  ^{\otimes n}\mathbf{x^{\prime}}\geq0, \label{PPTcons}%
\end{equation}
where
\begin{equation}
\mathbf{x^{\prime}}=\left(  x_{0},\overset{n}{\overbrace{\frac{x_{1}}%
{n},\ldots,\frac{x_{1}}{n}},}\ldots\overset{\binom{n}{k}}{\overbrace
{\frac{x_{k}}{\binom{n}{k}},\ldots,\frac{x_{k}}{\binom{n}{k}}}},\ldots
x_{n}\right)  ^{T}.
\end{equation}

For general $n$, it is not known if the PPT states
$\sigma_{\mathbf{x}}^{n}$ satisfying Eq.~(\ref{PPTcons}) are
separable. However, they are known to be equivalent for
$n=2$~\cite{VolWer}, in which case Eq.~(\ref{PPTcons}) simplifies
to
\begin{align}
1-2x_{1}  &  \geq0,\\
(d-1)-2dx_{0}+(2-d)x_{1}  &  \geq0,\\
(d-1)^{2}+4dx_{0}+2(d-1)x_{1}  &  \geq0,
\end{align}
where we have eliminated the dependent variable $x_{2}$. This means that, for
two copies ($n=2$), the state $\sigma_{\mathbf{x}}^{2}$ is the closest state
for the minimization of both $E_{P}(W_{\eta,d}^{\otimes2})$ and $E_{R}%
(W_{\eta,d}^{\otimes2})$. Let us compute the latter quantity.

Assuming the basis where the single-copy Werner state is diagonal, we may
write%
\begin{align}
S\left(  W_{\eta,d}^{\otimes n}||\sigma_{\mathbf{x}}^{n}\right)   &
=\sum_{i=0}^{n}y_{i}\mathrm{log}_{2}\left(  \frac{y_{i}}{x_{i}}\right)  ,\\
y_{i}  &  =\frac{\binom{n}{i}(1-\eta)^{n-i}(1+\eta)^{i}}{2^{n}}.
\end{align}
Therefore, for $n=2$ and in the region $\eta\leq-\frac{2}{d}$, we derive
\begin{gather}
E_{R}^{2}\left(  W_{\eta,d}\right)  :=\frac{E_{R}\left(  W_{\eta,d}^{\otimes
2}\right)  }{2}=\min_{x_{0},x_{1}}\left\{  \frac{(1-\eta)^{2}}{8}%
\mathrm{log}_{2}\frac{(1-\eta)^{2}}{4x_{0}}\right. \nonumber\\
+\frac{(1-\eta)(1-\eta)}{4}\mathrm{log}_{2}\frac{(1-\eta)(1-\eta)}{2x_{1}%
}\nonumber\\
\left.  +\frac{(1+\eta)^{2}}{8}\mathrm{log}_{2}\frac{(1+\eta)^{2}}{4\left(
1-x_{0}-x_{1}\right)  }\right\}  , \label{Lagrange}%
\end{gather}
where%
\begin{align}
1-2x_{1}  &  \geq0,\\
(d-1)-2dx_{0}+(2-d)x_{1}  &  \geq0,\\
(d-1)^{2}+4dx_{0}+2(d-1)x_{1}  &  \geq0,\\
x_{0}+x_{1}  &  \leq1.
\end{align}

\begin{figure}[ptbh]
\vspace{+0.2cm} \includegraphics[width=\columnwidth]{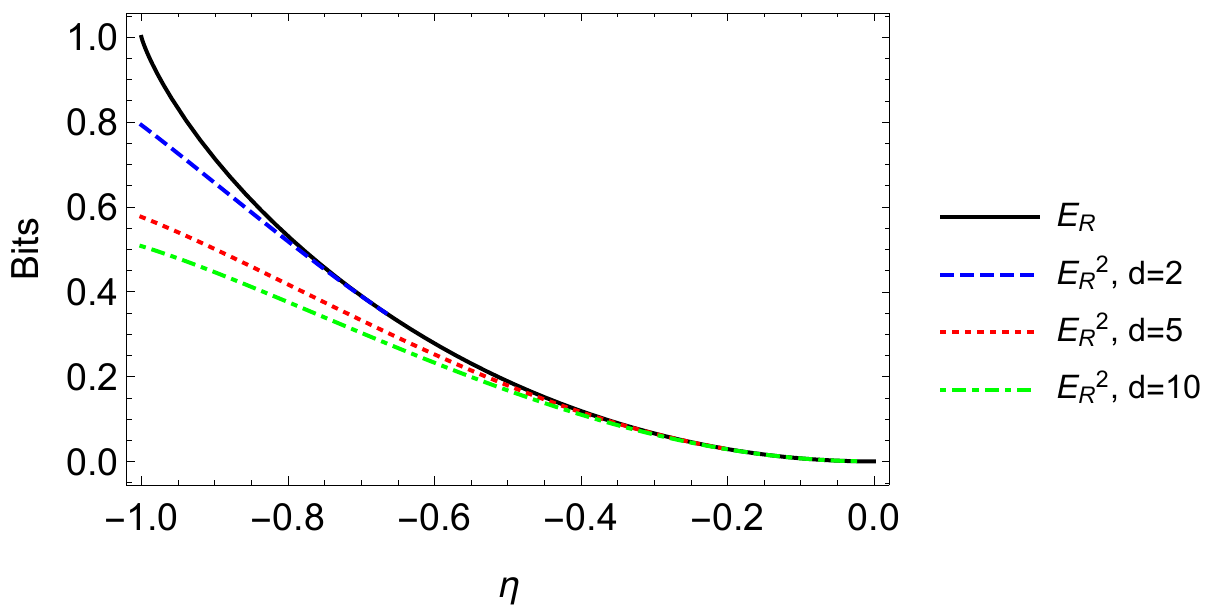}
\caption{Comparison between the one-copy REE $E_{R}$ and the two-copy REE
$E_{R}^{2}$ of a Werner state $W_{\eta,d}$, for varying dimension $d>2$. In
particular, we consider here $\eta\leq0$ which includes the subadditivity
region $\eta<-2/d$.}%
\label{compa}%
\end{figure}

We can use Lagrangian optimisation methods to solve this problem. Let us set%
\begin{align}
\theta &  :=d^{4}\left(  \eta^{2}+1\right)  ^{2}-4d^{3}\eta\left(  \eta
^{2}-3\right) \\
&  -4d^{2}\left(  \eta^{4}+3\eta^{2}-1\right)  +8d\eta\left(  \eta
^{2}-3\right)  +4\left(  \eta^{2}+1\right)  ^{2},\nonumber
\end{align}
then we compute
\begin{align}
x_{0}  &  =\frac{d^{2}\left(  \eta^{2}+1\right)  +\sqrt{\theta}-2d(\eta
-2)-2\eta^{2}-2}{8d(d+2)},\\
x_{1}  &  =-\frac{d^{2}\left(  \eta^{2}-3\right)  +\sqrt{\theta}-2d\eta
-2\eta^{2}+6}{4\left(  d^{2}-4\right)  }.
\end{align}
The comparison between the one-copy REE $E_{R}(W_{\eta,d})$ of
Eq.~(\ref{onecopyREE}) and the two-copy REE $E_{R}^{2}\left(  W_{\eta
,d}\right)  $ of Eq.~(\ref{Lagrange}) is shown in Fig.~\ref{compa}. While
$E_{R}(W_{\eta,d})$ does not depend on the dimension $d$, we see that the
two-copy REE considerably decreases for increasing $d>2$.

\section{Two-way assisted capacities of the Holevo-Werner
channels\label{Capacities}}

\subsection{Weak converse bounds based on the relative entropy distances}

We now combine the results in the previous section with the methods of
Ref.~\cite{Stretching} to bound the two-way capacities of the HW channels.
According to Ref.~\cite{Stretching}, the secret-key capacity $K$ of a
teleportation covariant channel $\mathcal{E}$ is upper bounded by the
regularised REE of its Choi Matrix $\chi_{\mathcal{E}}$, i.e.,
\begin{equation}
K\left(  \mathcal{E}\right)  \leq E_{R}^{\infty}\left(  \chi_{\mathcal{E}%
}\right)  . \label{upper}%
\end{equation}
Therefore, for an HW channel $\mathcal{W}_{\eta,d}$, we may write the upper
bound%
\begin{equation}
K(\mathcal{W}_{\eta,d})\leq E_{R}^{\infty}\left(  W_{\eta,d}\right)  ,
\end{equation}
by using its corresponding Werner state $W_{\eta,d}$. From the previous
section, we have that, for $\eta<-2/d$ we may write the following strict
inequality%
\begin{equation}
K(\mathcal{W}_{\eta,d})\leq E_{R}^{2}\left(  W_{\eta,d}\right)  <E_{R}\left(
W_{\eta,d}\right)  ,
\end{equation}
so that the one-copy (single-letter) REE bound is strictly loose. This shows
that the regularised REE is needed to tightly bound (and possibly establish)
the secret-key capacity of an HW channel. As shown in Fig.~\ref{compa}, the
improvement of $E_{R}^{2}$ over $E_{R}$ is better and better for increasing
dimension $d$.

Let us now consider the two-way quantum capacity $Q_{2}$, which is also known
to be equal to the channel's two-way entanglement distribution capacity
$D_{2}$. In Appendix~\ref{APPprova}, we provide a general proof of the following.

\begin{lemma}
[Channel's RPPT bound]For a teleportation covariant channel $\mathcal{E}$, we
may write
\begin{equation}
Q_{2}(\mathcal{E})\leq E_{P}^{\infty}\left(  \chi_{\mathcal{E}}\right)
,\label{RPPTbound}%
\end{equation}
where the Choi matrix $\chi_{\mathcal{E}}$ and the RPPT $E_{P}^{\infty}$ are
meant to be asymptotic if $\mathcal{E}$ is a continuous-variable channel. In
particular, $E_{P}^{\infty}\left(  \chi_{\mathcal{E}}\right)  $ becomes the
regularisation of
\begin{equation}
E_{P}\left(  \chi_{\mathcal{E}}\right)  :=\inf_{\sigma^{\mu}}\underset
{\mu\rightarrow+\infty}{\lim\inf}S(\chi_{\mathcal{E}}^{\mu}||\sigma^{\mu
}),\label{asyBBB}%
\end{equation}
where: $\chi_{\mathcal{E}}^{\mu}:=\mathcal{I}\otimes\mathcal{E}(\Phi^{\mu})$
is defined on a two-mode squeezed vacuum state $\Phi^{\mu}$ with energy $\mu$,
and $\sigma^{\mu}$ is a sequence of PPT states converging in trace norm, i.e.,
such that $\left\Vert \sigma^{\mu}-\sigma\right\Vert \overset{\mu}%
{\rightarrow}0$ for some PPT state $\sigma$.
\end{lemma}

\begin{figure}[ptb]
\vspace{+0.1cm} \includegraphics[width=\columnwidth]{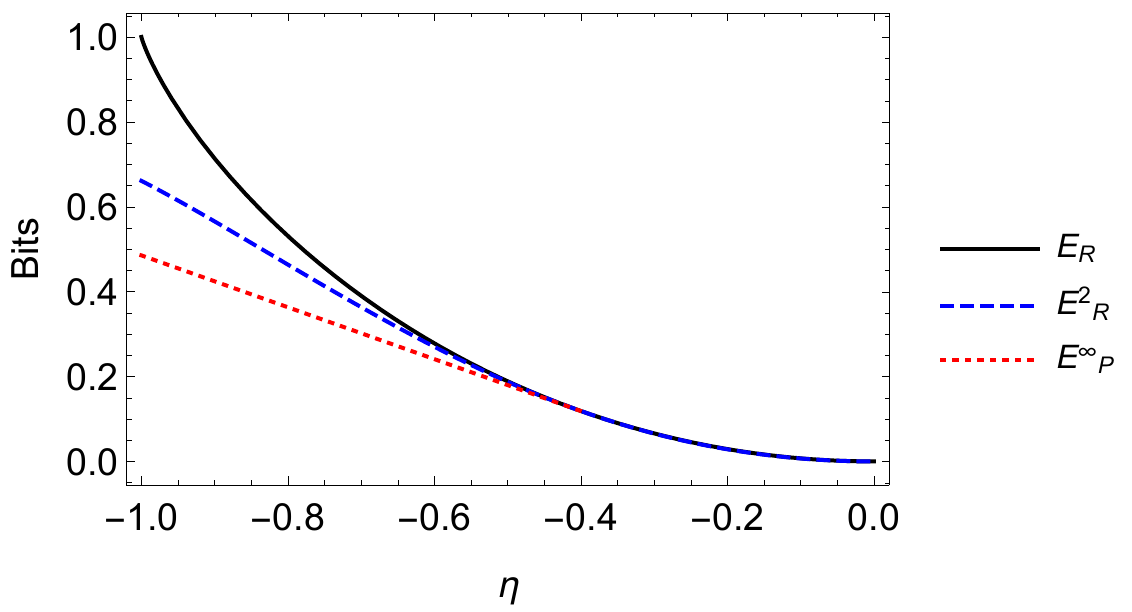}
\caption{Weak converse upper bounds for the two-way quantum capacity $Q_{2}$
of the HW channel $\mathcal{W}_{\eta,5}$ (dimension $d=5$). We compare the
one-copy REE bound $E_{R}(=E_{P})$, the two-copy REE bound $E_{R}^{2}\left(
=E_{P}^{2}\right)  $, and the regularised RPPT bound $E_{P}^{\infty}$, which
is the tightest. Note that $E_{R}$ and $E_{R}^{2}$ also bound the secret-key
capacity $K$ of the channel.}%
\label{queuetwocomp}%
\end{figure}

By applying the bound of Eq.~(\ref{RPPTbound}) to an HW channel $\mathcal{W}%
_{\eta,d}$, we may write
\begin{equation}
Q_{2}(\mathcal{W}_{\eta,d})\leq E_{P}^{\infty}\left(  W_{\eta,d}\right)  ,
\label{bb1}%
\end{equation}
where the right hand side is computed as in Eq.~(\ref{PPTres}). Of course we
may also write
\begin{equation}
Q_{2}(\mathcal{W}_{\eta,d})\leq E_{R}^{2}\left(  W_{\eta,d}\right)  \leq
E_{R}\left(  W_{\eta,d}\right)  =E_{P}\left(  W_{\eta,d}\right)  . \label{bb2}%
\end{equation}
The bounds in Eqs.~(\ref{bb1}) and~(\ref{bb2}) are shown and compared in
Fig.~\ref{queuetwocomp} for a HW channel in dimension $d=5$%
.

\subsection{Weak converse bounds based on the squashed entanglement}

Whilst the relative entropy distances provide useful upper bounds, we may also
consider other functionals. In particular, we may consider the squashed
entanglement. For an arbitrary bipartite state $\rho_{AB}$, this is defined
as~\cite{squash,HayashiINTRO}
\begin{equation}
E_{sq}(\rho_{AB}):=\frac{1}{2}\min_{\rho_{ABE}^{\prime}\in\Omega_{AB}%
}S(A:B|E),
\end{equation}
where $\Omega_{AB}$ is the set of density matrices $\rho_{ABE}^{\prime}$
satisfying $\mathrm{Tr}_{E}(\rho_{ABE}^{\prime})=\rho_{AB}$, and $S(A:B|E)$ is
the conditional mutual information
\begin{equation}
S(A:B|E):=S(\rho_{AE}^{\prime})+S(\rho_{BE}^{\prime})-S(\rho_{E})-S(\rho
_{ABE}),
\end{equation}
with $S(...)$ being the Von Neumann entropy~\cite{NielsenChuang}.

The squashed entanglement can be combined with teleportation
stretching~\cite{Stretching} to provide a single-letter bound to the
secret-key capacity. In fact, it satisfies all the required conditions. It
normalises, so that $E_{sq}(\phi_{m})\geq mR_{m}$ for a private state
$\phi_{m}$ with $mR_{m}$ private bits~\cite{squash}. It is continuous, and
monotonic under LOCC~\cite{squash}. Furthermore, it is additive over
tensor-product states, which means that there is no need to regularize over
many copies. For a teleportation covariant dicrete-variable channel
$\mathcal{E}$, we may therefore write
\begin{equation}
K(\mathcal{E})\leq E_{sq}(\chi_{\mathcal{E}}).\label{Esq}%
\end{equation}
This is a direct consequence of Proposition 6 of Ref.~\cite{Stretching},
according to which we may write
\begin{equation}
K(\mathcal{E})=K(\chi_{\mathcal{E}}),
\end{equation}
where the latter is the distillable key of the Choi matrix $\chi_{\mathcal{E}%
}$. Then, using Ref.~\cite{squash}, we may write $K(\chi_{\mathcal{E}})\leq
E_{sq}(\chi_{\mathcal{E}})$, which leads to Eq.~(\ref{Esq})~\cite{ProvaCV}.

However, there is some difficulty in optimizing over $\rho_{ABE}^{\prime}$
such that $\mathrm{Tr}_{E}(\rho_{ABE}^{\prime})=\chi_{\mathcal{E}}$, since the
dimension of the environment system $E$ is generally unbounded. In order to
provide an analytical upper bound, we simply choose the purification
$\tilde{\chi}_{\mathcal{E}}$ of $\chi_{\mathcal{E}}$. In the case of an HW
channel $\mathcal{E}=\mathcal{W}_{\eta,d}$, we have $\chi_{\mathcal{E}%
}=W_{\eta,d}$ and we may write%
\begin{align}
K(\mathcal{W}_{\eta,d})  &  \leq E_{sq}(W_{\eta,d})\nonumber\\
&  \leq\tilde{E}_{sq}(W_{\eta,d}):=\frac{1}{2}S(A:B|E)_{\tilde{W}_{\eta,d}%
}\nonumber\\
&  =\log_{2}d+\frac{1+\eta}{4}\log_{2}\frac{1+\eta}{d(d+1)}\nonumber\\
&  +\frac{1-\eta}{4}\log_{2}\frac{1-\eta}{d(d-1)}, \label{sqKKK}%
\end{align}
which is positive only if $\eta\leq0$.

We can find a further upper bound by exploiting the convexity property of the
squashed entanglement. First note that
\begin{align}
W_{\eta,d}  &  =\frac{\left(  d-\eta\right)  \mathbb{I}+\left(  d\eta
-1\right)  \mathbb{F}}{d^{3}-d}\nonumber\\
&  =\left(  1+\eta\right)  W_{0,d}+\left(  -\eta\right)  W_{-1,d},
\end{align}
which means that for $-1\leq\eta\leq0$ the state $W_{\eta,d}$ can be written
as a convex combination of the separable state $W_{0,d}$ and the extremal
Werner state $W_{-1,d}$. Second, note that we have $E_{sq}\left(
W_{0,d}\right)  =0$ (since it is a separable state) and, for the extremal
state, we may write~\cite{entanglementantisymmetricstate}%

\begin{equation}
E_{sq}\left(  W_{-1,d}\right)  \leq%
\begin{cases}
\log_{2}\left(  \frac{d+2}{d}\right)   & \text{if }d\text{ even,}\\
\frac{1}{2}\log_{2}\left(  \frac{d+3}{d-1}\right)   & \text{if }d\text{
uneven.}%
\end{cases}
\end{equation}
Using the convexity property of the squashed entanglement~\cite{squash}
\begin{align}
&  E_{sq}\left[  p\rho_{1}+(1-p)\rho_{2}\right]  \nonumber\\
&  \leq pE_{sq}\left(  \rho_{1}\right)  +(1-p)E_{sq}\left(  \rho_{2}\right)  ,
\end{align}
we find that
\begin{equation}
K(\mathcal{W}_{\eta,d})\leq E_{sq}\left(  W_{\eta,d}\right)  \leq E_{sq}%
^{\ast}\left(  W_{\eta,d}\right)  ,
\end{equation}
where we define
\begin{equation}
E_{sq}^{\ast}\left(  W_{\eta,d}\right)  =%
\begin{cases}
-\eta\log_{2}\left(  \frac{d+2}{d}\right)   & \text{if }d\text{ even,}\\
-\frac{\eta}{2}\log_{2}\left(  \frac{d+3}{d-1}\right)   & \text{if }d\text{
uneven,}%
\end{cases}
\label{convexBBB}%
\end{equation}
for $-1\leq\eta\leq0$ and zero otherwise.

\begin{figure}[ptb]
\begin{center}
\vspace{-1.2cm}
\includegraphics[width=0.5\textwidth]{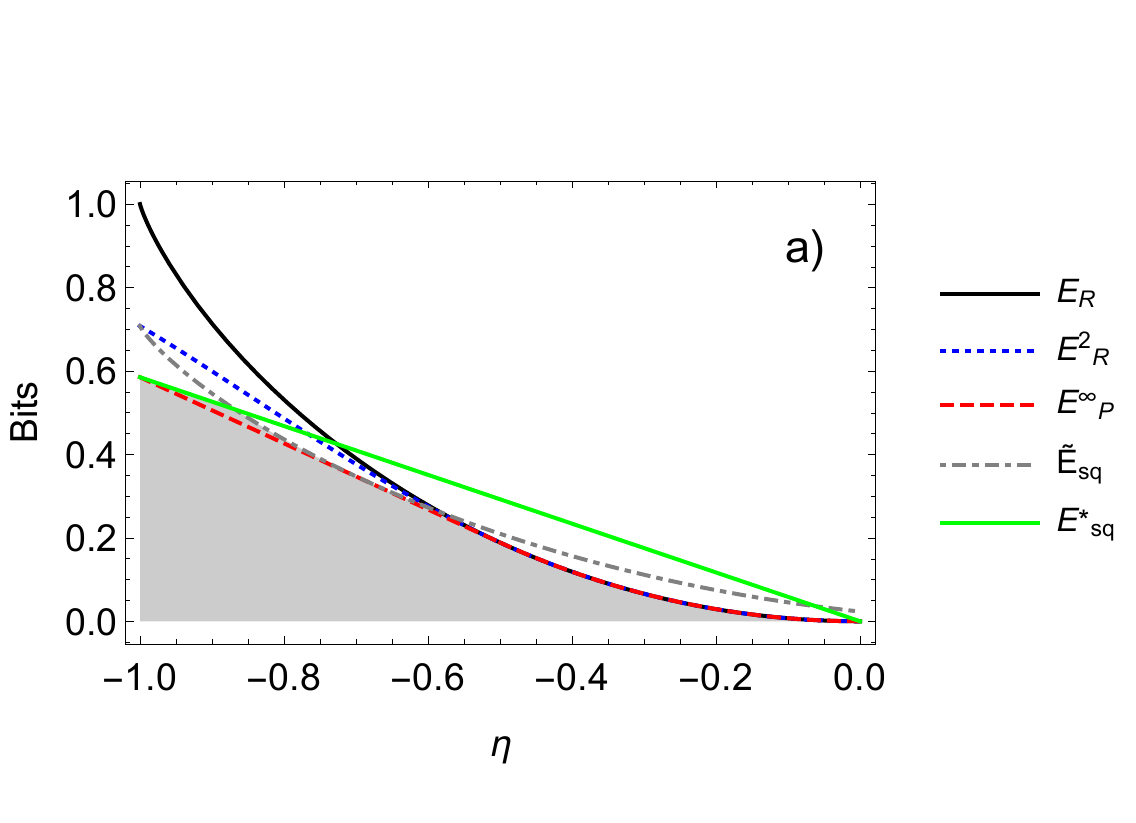}
\end{center}
\begin{center}
\vspace{-2.0cm}
\includegraphics[width=0.5\textwidth]{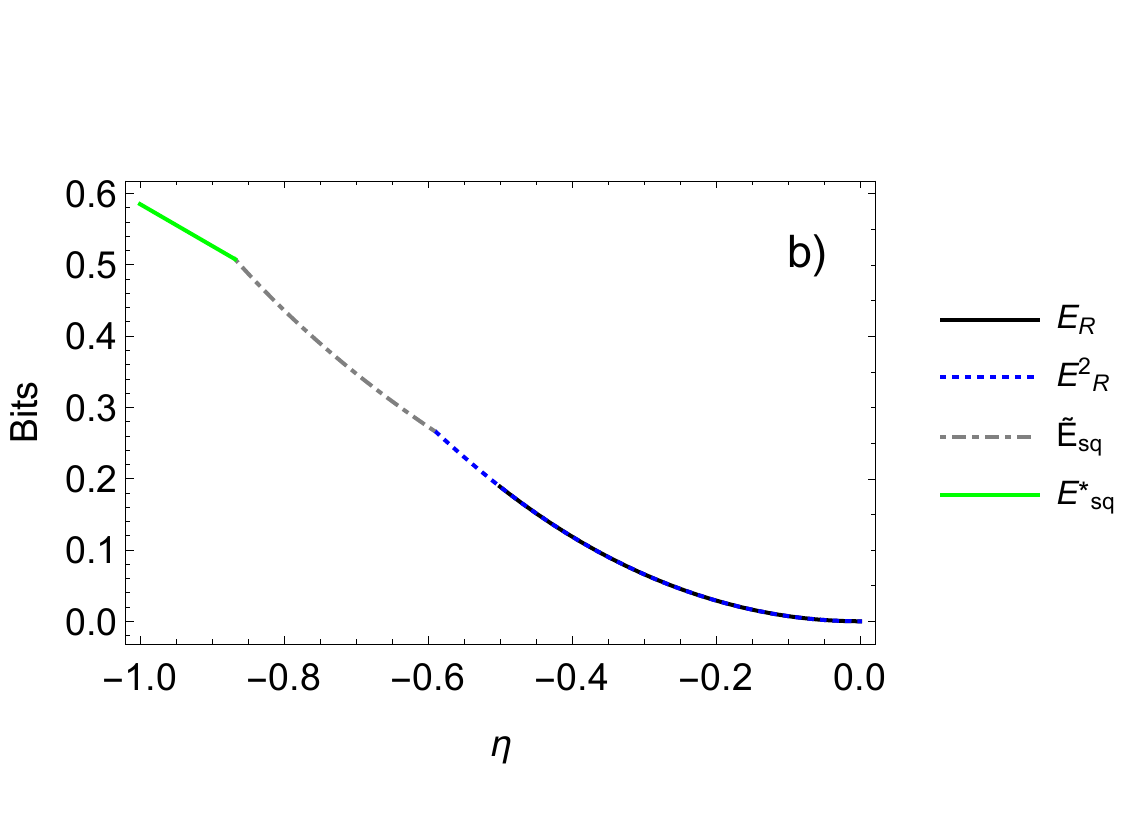}
\end{center} \vspace{-0.6cm} \caption{Comparison of the capacity
bounds for the HW channel $\mathcal{W}_{\eta,4}$. (a) The
regularised RPPT bound $E_{P}^{\infty}$ is the lowest (red-dashed)
curve and bounds the two-way quantum capacity $Q_{2}$ of the
channel. The secret-key capacity of the channel $K$ is in the gray
area. Depending on the value of $\eta$, this is upper-bounded by
the two-copy REE bound $E_{R}^{2}(=E_{P}^{2})$ (better than
$E_{R}(=E_{P})$) or by the squashed entanglement bounds
$\tilde{E}_{sq}$ and $E_{sq}^{\ast}$. We see that $\tilde{E}_{sq}$
coincides with $E_{R}^{2}$ for $\eta=-1$. (b) We show the
competing upper bounds for the secret-key capacity $K$ of the HW
channel $\mathcal{W}_{\eta,4}$, explicitly drawing which bound is
better at which value of $\eta$. We see that the squashed
entanglement bounds perform better at lower $\eta$, while the REE
bounds are better for
higher $\eta$. }%
\label{squashPIC}%
\end{figure}

These bounds are compared in Fig.~\ref{squashPIC} for the case of an HW
channel with dimension $d=4$. We can see that one bound is better than another
depending on the value of $\eta$. In particular, the secret-key capacity is in
the gray area of Fig.~\ref{squashPIC}(a) or, equivalently, below the
composition of bounds shown in Fig.~\ref{squashPIC}(b).


\section{Holevo-Werner Repeater Chains and Quantum Networks\label{SECnet}}

\subsection{Repeater chains}

\begin{figure}[ptbh]
\begin{center}
\vspace{-0.9cm} \includegraphics[width=0.5\textwidth]{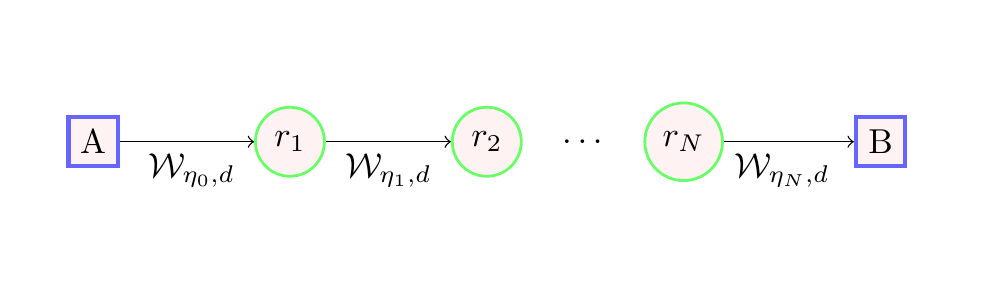}
\vspace{-1.2cm}
\end{center}
\caption{Alice (A) and Bob (B) are connected by $N$ quantum repeaters $r_{1}%
$,\ldots, $r_{N}$ in a linear chain; each connection (edge) in the chain is a
$d$ dimension HW channel with a generally-different parameter $\eta_{i}$.}%
\label{tikzchain}%
\end{figure}

In this section, we apply the results of Ref.~\cite{ref1} to bound the
end-to-end capacities of quantum networks in which the edges between nodes are
HW channels. First, we consider the simplest multi-hop quantum network which
consists of a linear chain of $N$ repeaters between the two end-parties. Such
a set up is depicted in Fig.~\ref{tikzchain}.

For a linear chain of $N$ quantum repeaters, whose $N+1$ connecting channels
$\{\mathcal{E}_{i}\}_{i=0}^{N}$ are teleportation covariant, we have that the
secret capacity $K$ of the chain and its two-way quantum capacity $Q_{2}$ are
bounded by~\cite{ref1}
\begin{align}
Q_{2} &  \leq K\leq\min_{i}E_{R}^{\infty}\left(  \chi_{\mathcal{E}_{i}%
}\right)  \nonumber\\
&  \leq\min_{i}E_{R}^{2}\left(  \chi_{\mathcal{E}_{i}}\right)  \leq\min
_{i}E_{R}\left(  \chi_{\mathcal{E}_{i}}\right)  ,
\end{align}
with $\chi_{\mathcal{E}_{i}}$ the Choi matrix of the $i^{\text{th}}$ channel.
Similarly, we may use the squashed entanglement and write~\cite{ref1}%
\begin{align}
Q_{2}  & \leq K\leq\min_{i}E_{sq}\left(  \chi_{\mathcal{E}_{i}}\right)
\nonumber\\
& \leq\min\{\min_{i}\tilde{E}_{sq}\left(  \chi_{\mathcal{E}_{i}}\right)
,\min_{i}E_{sq}^{\ast}\left(  \chi_{\mathcal{E}_{i}}\right)  \}.
\end{align}
In general, we may write
\begin{equation}
Q_{2}\leq K\leq\min_{E}\min_{i}E\left(  \chi_{\mathcal{E}_{i}}\right)  ,
\end{equation}
where the bound is also minimized over the type of entanglement measure. In
particular, we may consider the \textquotedblleft ideal\textquotedblright\ set
$E\in\{E_{R}^{\infty},E_{sq}\}$ or the \textquotedblleft
computable\textquotedblright\ one $E\in\{E_{R}^{2}\leq E_{R},\tilde{E}%
_{sq},E_{sq}^{\ast}\}$. Then, if the task of the parties is to transmit qubits
(or distill ebits), we may use the regularised RPPT and write~\cite{ref1}%
\begin{equation}
Q_{2}=D_{2}\leq\min_{i}E_{P}^{\infty}\left(  \chi_{\mathcal{E}_{i}}\right)  .
\end{equation}

Let us apply these results to a linear repeater chain connected by $N+1$
iso-dimensional HW channels $\left\{  \mathcal{W}_{\eta_{i},d}\right\}
=\left\{  \mathcal{W}_{\eta_{0},d},\ldots,\mathcal{W}_{\eta_{N},d}\right\}  $,
i.e., with the same dimension $d$ but generally different $\eta$'s. We may
simplify the previous bounds ($E_{R}$, $E_{R}^{2}$, $\tilde{E}_{sq}$,
$E_{sq}^{\ast}$, and $E_{P}^{\infty}$) by exploiting the fact that they are
monotonically decreasing in $\eta$, so that the maximum value $\eta
_{\text{max}}:=\max\left\{  \eta_{i}\right\}  $ determines the bottleneck of
the chain, i.e., $\min_{i}E=E(W_{\eta_{\text{max}},d})$. In particular, for
$\eta_{\text{max}}\geq0$, we certainly have $Q_{2}=D_{2}=K=0$ because
$E_{R}(W_{\eta_{\text{max}}\geq0,d})=0$ from Eq.~(\ref{onecopyREE}). By
contrast, if $\eta_{\text{max}}\leq0$, then we may write the following bounds
for the secret-key capacity and two-way quantum capacity of the repeater chain%
\begin{align}
K\left(  \left\{  \mathcal{W}_{\eta_{i},d}\right\}  \right)    & \leq\min
_{E}E\left(  W_{\eta_{\text{max}},d}\right)  ,\label{Kcompat}\\
Q_{2}\left(  \left\{  \mathcal{W}_{\eta_{i},d}\right\}  \right)    & \leq
E_{P}^{\infty}\left(  W_{\eta_{\text{max}},d}\right)  .\label{Qcompact}%
\end{align}
In Eq.~(\ref{Kcompat}), the optimal entanglement measure $E$ can be computed
from the set $\{E_{R}^{2}\leq E_{R},\tilde{E}_{sq},E_{sq}^{\ast}\}$, where
$E_{R}$ is given in Eq.~(\ref{onecopyREE}), $E_{R}^{2}$ in Eq.~(\ref{Lagrange}%
), $\tilde{E}_{sq}$ in Eq.~(\ref{sqKKK}), $E_{sq}^{\ast}$ in
Eq.~(\ref{convexBBB}). In Eq.~(\ref{Qcompact}), we compute $E_{P}^{\infty}$
from Eq.~(\ref{PPTres}).

\subsection{Single-path routing in quantum networks}

We may then extend the results to an arbitrary quantum network, where there
exist many possible paths between the two end-parties, Alice and Bob. Assuming
single-path routing, a single chain of repeaters is used for each use of the
network and this may differ from use to use. For a network connected by
teleportation covariant channels, we may bound the single-path secret-key
capacity of the network as~\cite{ref1}%
\begin{equation}
K\leq\min_{C}E(C),~E(C):=\max_{\mathcal{E}\in\tilde{C}}E(\chi_{\mathcal{E}%
}),\label{cutset}%
\end{equation}
where $E$ is a suitable entanglement measure, here to be optimized in
$\{E_{R},E_{sq}\}$~\cite{remark}, and $\tilde{C}$ is a \textquotedblleft
cut-set\textquotedblright\ associated with the cut~\cite{Slepian,netflow}.

The cut-set $\tilde{C}$ can be described as a set of channels such that, if
those channels were removed by the cut, then the network would be
bi-partitioned, with Alice and Bob in separate sets of nodes. Therefore the
meaning of Eq.~(\ref{cutset}) is that: (i)~we perform an arbitrary cut $C$ of
the network; (ii)~we consider the channels $\mathcal{E}$ in the cut-set
$\tilde{C}$; (iii)~we compute the entanglement measure $E$ of their Choi
matrices $\chi_{\mathcal{E}}$; (iv)~we take the maximum so as to compute
$E(C)$; (v)~we finally minimize over all the possible Alice-Bob cuts $C$ of
the network.

\begin{figure}[ptbh]
\begin{center}
\vspace{-0.8cm} \includegraphics[width=0.5\textwidth]{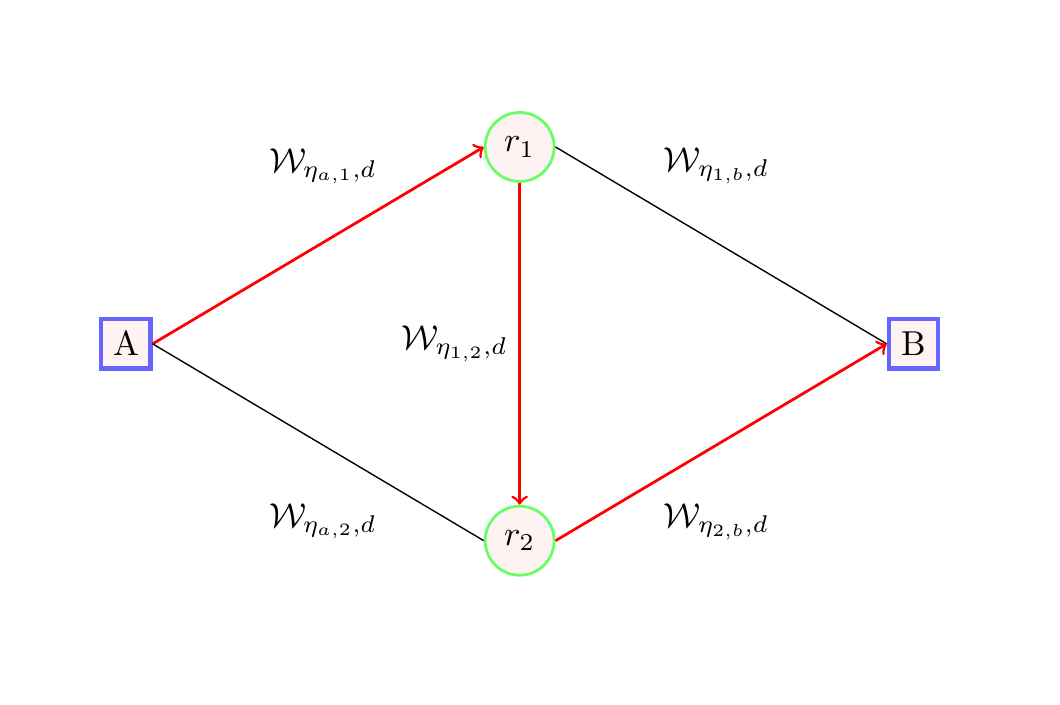}
\vspace{-1.4cm}
\end{center}
\caption{Alice (A) and Bob (B) as end-nodes of a diamond network connected by
iso-dimensional HW\ channels with generally-different expectation parameters
$\eta$. In red we show a possible path between the end-nodes.}%
\label{DiamondNet}%
\end{figure}

In the case of a quantum network connected by HW channels, we may the
following bound for the single-path secret-key capacity
\begin{equation}
K\leq\min_{C}\max_{\mathcal{W}_{\eta,d}\in\tilde{C}}E(W_{\eta,d}).
\end{equation}
If the HW channels are iso-dimensional (as in the example of
Fig.~\ref{DiamondNet}), then we may simplify the previous bound into the
following%
\begin{equation}
K\leq\min_{C}E(W_{\eta_{\text{min}(C)},d}),
\end{equation}
where $\eta_{\text{min}(C)}$ is the smallest expectation parameter belonging
to the cut-set $\tilde{C}$. In particular, we may also miminize $E$ over
$\{E_{R},\tilde{E}_{sq},E_{sq}^{\ast}\}$ by computing $E_{R}$ as in
Eq.~(\ref{onecopyREE}), $\tilde{E}_{sq}$ as in Eq.~(\ref{sqKKK}), and
$E_{sq}^{\ast}$ as in Eq.~(\ref{convexBBB}).

\subsection{Multi-path routing in quantum networks}

\begin{figure}[ptbh]
\begin{center}
\vspace{-0.4cm} \includegraphics[width=0.5\textwidth]{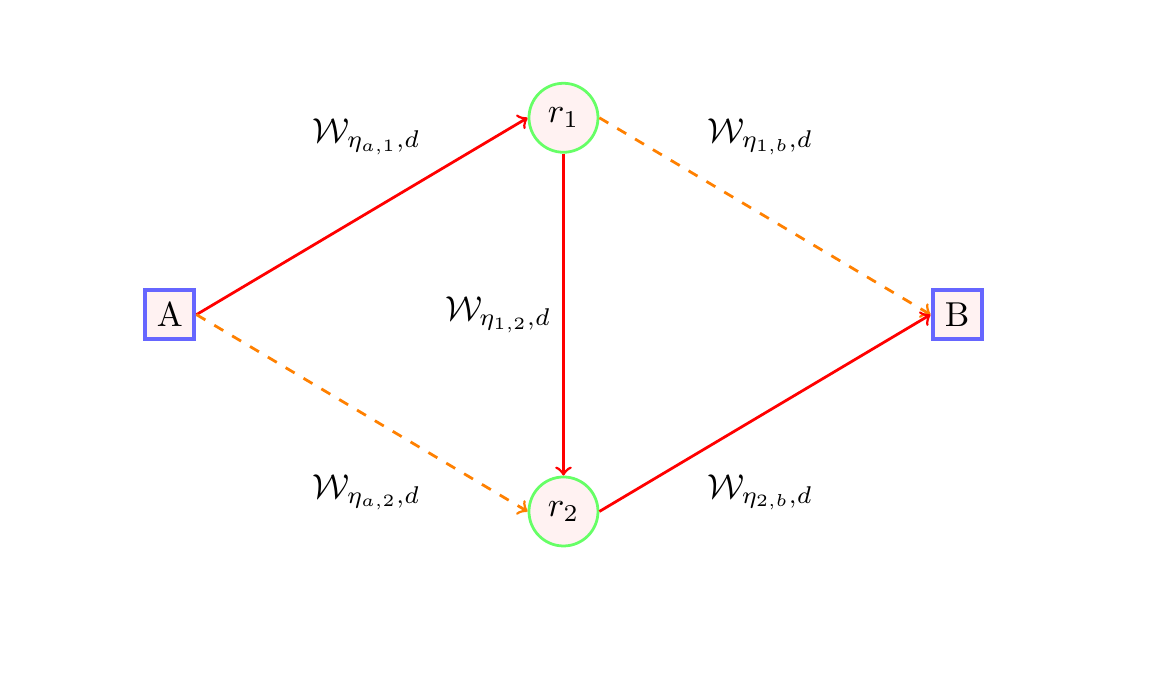}
\vspace{-1.5cm}
\end{center}
\caption{Example of multipath routing in a diamond network (with
iso-dimensional HW channels). With respect to Fig.~\ref{DiamondNet} all the
channels are used in a single use of the network (flooding protocol). Dashed
lines represent the advantage over the previous single-path routing protocol.}%
\label{DiaFlood}%
\end{figure}

Finally we may also consider multipath routing. In this case, each use of the
network corresponds to a simultaneous use of all the channels, allowing for
simultaneous pathways between Alice and Bob (e.g., see Fig.~\ref{DiaFlood}).
This is also known as a flooding protocol~\cite{flooding} and represents a
crucial requirement in order to extend the max-flow/min-cut
theorem~\cite{Harris,Ford,ShannonFLOW} to the quantum setting~\cite{ref1}.

For a network connected by teleportation-covariant channels, the multi-path
secret-key capacity $K^{\text{m}}\geq K$ is bounded as~\cite{ref1}%
\begin{align}
K^{\text{m}} &  \leq\min_{C}\Sigma^{\infty}(C)\leq\cdots\leq\min_{C}\Sigma
^{r}(C)\nonumber\\
&  \leq\cdots\leq\min_{C}\Sigma^{1}(C),
\end{align}
where, for any integer $r=1,\cdots,\infty$,%
\begin{equation}
\Sigma^{r}(C):=\sum_{\mathcal{E}\in\tilde{C}}E^{r}(\chi_{\mathcal{E}}).
\end{equation}
and $E^{r}$ is a suitable $r$-copy entanglement measure. In particular, we may
optimize over the multi-copy REE $E^{r}=E_{R}^{r}$ or the squashed
entanglement $E^{r}=E_{sq}$ (the latter being additive). For the multipath
two-way quantum capacity, we may correspondingly write%
\begin{equation}
Q_{2}^{\text{m}}\leq\min_{C}\Sigma_{P}^{\infty}(C)\leq\cdots\leq\min_{C}%
\Sigma_{P}^{1}(C),
\end{equation}
where
\begin{equation}
\Sigma_{P}^{r}(C):=\sum_{\mathcal{E}\in\tilde{C}}E_{P}^{r}(\chi_{\mathcal{E}%
}),
\end{equation}
and $E_{P}^{r}$ is the $r$-copy RPPT.

For a network connected by HW channels $\mathcal{W}_{\eta,d}$, we may specify
the previous bounds to one- and two-copy REE, so that we may write%
\begin{equation}
K^{\text{m}}\leq\min_{C}\sum_{\mathcal{W}_{\eta,d}\in\tilde{C}}E_{R}%
^{2}(W_{\eta,d})\leq\min_{C}\sum_{\mathcal{W}_{\eta,d}\in\tilde{C}}%
E_{R}(W_{\eta,d}),\label{Kemme}%
\end{equation}
where $E_{R}$ is in Eq.~(\ref{onecopyREE}), and $E_{R}^{2}$ in
Eq.~(\ref{Lagrange}). The first bound in Eq.~(\ref{Kemme}) is
certainly tighter than the second one if the channels have
$\eta<-2/d$.\ More generally, we write
\begin{equation}
K^{\text{m}}\leq\min_{E}\min_{C}\sum_{\mathcal{W}_{\eta,d}\in\tilde{C}%
}E(W_{\eta,d}),
\end{equation}
where $E$ is minimized in the computable set $\{E_{R}^{2}\leq E_{R},\tilde
{E}_{sq},E_{sq}^{\ast}\}$. Finally,\ we may write
\begin{equation}
Q_{2}^{\text{m}}\leq\min_{C}\sum_{\mathcal{W}_{\eta,d}\in\tilde{C}}%
E_{P}^{\infty}(W_{\eta,d})\leq\min_{C}\sum_{\mathcal{W}_{\eta,d}\in\tilde{C}%
}E_{P}(W_{\eta,d}),\label{multiQ2}%
\end{equation}
where $E_{P}$ is in Eq.~(\ref{onecopyREE}) and $E_{P}^{\infty}$ in
Eq.~(\ref{PPTres}). The first bound in Eq.~(\ref{multiQ2}) is computable from
the regularised RPPT in Eq.~(\ref{PPTres}) and is certainly strictly tigther
than the second bound if the channels have $\eta<-2/d$.

\section{Conclusions\label{Werconclu}}

In this work we have considered quantum and private communication over the
class of (teleportation-covariant) Holevo-Werner channels. We have computed
suitable upper bounds for their two-way assisted capacities in terms of
relative entropy distances, i.e., the relative entropy of entanglement (REE)
and its variant with respect to PPT\ states (RPPT), and also in terms of the
squashed entanglement (using the identity isometry and then the convexity
property).

We have shown that there is a general competing behaviour between these
bounds, so that an optimization over the entanglement measure is in order.
These calculations were done not only for point-to-point communication, but
also for chains of quantum repeaters and, more generally, quantum networks
under different types of routings.

In all cases, we have also pointed out the subadditivity behaviour of the REE
and RPPT bounds, so that their two-copy and regularised versions perform
strictly better than their simpler one-copy expressions, under suitable
conditions of the parameters. From this point of view, our paper clearly shows
how the subadditivity properties of the Werner states can be fully mapped to
the corresponding Holevo-Werner channels in configurations of adaptive quantum
and private communication.

\smallskip


\textit{Acknowledgements}.--This work has been supported by the EPSRC via the
`UK Quantum Communications Hub' (EP/M013472/1) and by the Innovation Fund
Denmark (Qubiz project). The authors would like to thank David Elkouss for feedback.

\bigskip

\appendix

\section{Proof of the RPPT bound in Lemma~1 at any dimension\label{APPprova}}

\subsection{Discrete-variable channels}

To get the result for finite dimension, we may apply an heuristic argument of
reduction into entanglement distillation~\cite{B2main} (suitably extended from
Pauli channels to teleportation-covariant channels). This gives $Q_{2}%
(\mathcal{E})=D_{2}(\chi_{\mathcal{E}})$, where the latter is the two-way
distillability of the Choi matrix $\chi_{\mathcal{E}}$. Then, we may use the
fact that $D_{2}(\chi_{\mathcal{E}})\leq E_{P}^{\infty}\left(  \chi
_{\mathcal{E}}\right)  $~\cite[Sec. 8.10]{HayashiINTRO}, therefore deriving
the bound in Eq.~(\ref{RPPTbound}) for discrete variable channels.

This bound can be proven more rigorously (and also extended to bosonic
channels), by resorting to teleportation stretching~\cite{Stretching}, where
the $n$-use output of a quantum protocol $\rho^{n}$ is directly expressed in
terms of the resource states ($\chi_{\mathcal{E}}^{\otimes n}$) via a single
but complicated trace-preserving LOCC $\Lambda$, i.e.,
\begin{equation}
\rho^{n}=\Lambda\left(  \chi_{\mathcal{E}}^{\otimes n}\right)  .
\label{teleST}%
\end{equation}

Recall that, for any channel $\mathcal{E}$, we may consider an adaptive
entanglement-distillation protocol $\mathcal{P}$ such that, after $n$ uses,
Alice and Bob share an output state $\rho^{n}$ satisfying the trace-distance
condition $||\rho^{n}-\Phi_{2}^{\otimes nR_{n}}||_{1}\leq\varepsilon$, where
$\Phi_{2}^{\otimes nR_{n}}$ are $nR_{n}$ ebits. By taking the limit in $n$ and
optimizing over $\mathcal{P}$, we write
\begin{equation}
Q_{2}(\mathcal{E})=D_{2}(\mathcal{E})=\sup_{\mathcal{P}}\lim_{n\rightarrow
\infty}R_{n}.
\end{equation}
Then recall the asymptotic continuity: For any pair of finite-dimensional
bipartite states, $\rho$ and $\sigma$, such that $||\rho-\sigma||_{1}%
\leq\varepsilon$, we may write $|E_{P}(\rho)-E_{P}(\sigma)|\leq f(\varepsilon
,d)$, where~\cite{Horos,Donald,Winter}
\begin{equation}
f(\varepsilon,d):=\frac{\varepsilon}{2}\mathrm{log}_{2}d+\left(
1+\frac{\varepsilon}{2}\right)  H_{2}\left(  \frac{\varepsilon}{2+\varepsilon
}\right)  ,
\end{equation}
with $H_{2}$ being the binary Shannon entropy~\cite{Cover&Thomas} and $d$ the
smaller of the two subsystems' dimensions. For any finite $d$, this function
$f$ disappears as $\varepsilon\rightarrow0$. Using this property and the
normalization~\cite{Rains} $E_{P}(\Phi_{2}^{\otimes nR_{n}})\geq nR_{n}$, we
may write
\begin{equation}
nR_{n}\leq E_{P}(\Phi_{2}^{\otimes nR_{n}})\leq E_{P}(\rho^{n})+f(\varepsilon
,d^{nR_{n}}). \label{eqprev}%
\end{equation}

Next step is to apply teleportation stretching to reduce the output state
$\rho^{n}$. For an adaptive protocol over a finite-dimensional
teleportation-covariant channel, we may write Eq.~(\ref{teleST}) where
$\chi_{\mathcal{E}}$ is the channel's Choi matrix and $\Lambda$ is a
trace-preserving LOCC~\cite{Stretching}. Because the RPPT is monotonic under
PPT\ operations, it is so under the more restrictive LOCCs as $\Lambda$.
Therefore, we may write $E_{P}(\rho^{n})\leq E_{P}(\chi_{\mathcal{E}}^{\otimes
n})$ and Eq.~(\ref{eqprev}) becomes
\begin{gather}
nR_{n}\leq E_{P}\left(  \chi_{\mathcal{E}}^{\otimes n}\right)  +f(\varepsilon
,d^{nR_{n}})\\
=E_{P}\left(  \chi_{\mathcal{E}}^{\otimes n}\right)  +\frac{\varepsilon
nR_{n}}{2}\mathrm{log}_{2}d\nonumber\\
+\left(  1+\frac{\varepsilon}{2}\right)  H_{2}\left(  \frac{\varepsilon
}{2+\varepsilon}\right)  .
\end{gather}

By re-organizing the terms in the previous inequality, we may write%
\begin{equation}
R_{n}\leq\frac{E_{P}\left(  \chi_{\mathcal{E}}^{\otimes n}\right)  +\left(
1+\frac{\varepsilon}{2}\right)  H_{2}\left(  \frac{\varepsilon}{2+\varepsilon
}\right)  }{n\left(  1-\frac{\varepsilon}{2}\mathrm{log}_{2}d\right)  }.
\end{equation}
Taking the limit in $n$, we therefore get
\begin{equation}
\lim_{n\rightarrow\infty}R_{n}\leq\frac{E_{P}^{\infty}\left(  \chi
_{\mathcal{E}}\right)  }{1-\frac{\varepsilon}{2}\mathrm{log}_{2}d}.
\label{Toextend}%
\end{equation}
For $\varepsilon\rightarrow0$ (weak converse), we obtain $\lim_{n\rightarrow
\infty}R_{n}\leq E_{P}^{\infty}\left(  \chi_{\mathcal{E}}\right)  $ and the
optimization over the protocols $\mathcal{P}$ automatically leads to the upper
bound $Q_{2}(\mathcal{E})\leq E_{P}^{\infty}\left(  \chi_{\mathcal{E}}\right)
$ as promised in Eq.~(\ref{RPPTbound}).

\subsection{Continuous-variable channels}

Thanks to the latter derivation, we can extend the bound to
continuous-variable (bosonic) channels, for which the output state $\rho^{n}$
is infinite-dimensional. Following Ref.~\cite{Stretching}, we apply a
truncation LOCC $\mathbb{T}_{d}$ at the output of the protocol $\mathcal{P}$
so that $\rho^{n,d}=\mathbb{T}_{d}(\rho^{n})$ is a finite dimensional state,
epsilon-close to $nR_{n,d}$ ebits. We may then repeat the previous steps and
modify Eq.~(\ref{eqprev}) into
\begin{align}
nR_{n,d}  &  \leq E_{P}(\rho^{n,d})+f(\varepsilon,d^{nR_{n,d}})\\
&  \leq E_{P}(\rho^{n})+f(\varepsilon,d^{nR_{n,d}}), \label{Tocomb1}%
\end{align}
where we exploit the monotoniticy $E_{P}(\rho^{n,d})\leq E_{P}(\rho^{n})$ in
the second inequality.

Now we use the asymptotic stretching $\rho^{n}=\lim_{\mu}\Lambda
(\chi_{\mathcal{E}}^{\mu\otimes n})$ in terms of the quasi-Choi matrix
$\chi_{\mathcal{E}}^{\mu}:=\mathcal{I}\otimes\mathcal{E}(\Phi^{\mu})$, with
$\Phi^{\mu}$ being a two-mode squeezed vacuum state with energy $\mu$. More
precisely, we write
\begin{equation}
\left\Vert \rho^{n}-\Lambda(\chi_{\mathcal{E}}^{\mu\otimes n})\right\Vert \leq
n\varepsilon_{\mu,N},
\end{equation}
where $\varepsilon_{\mu,N}:=\left\Vert \mathcal{E}-\mathcal{E}^{\mu
}\right\Vert _{\diamond N}$ is the channel simulation error expressed in terms
of energy-constrained diamond distance between the channel $\mathcal{E}$ and
its teleportation simulation $\mathcal{E}^{\mu}$~\cite{Stretching}. For any
finite energy $N$ of the input alphabet, we have the bounded-uniform
convergence of the Braunstein-Kimble protocol, so that $\lim_{\mu}%
\varepsilon_{\mu,N}=0$. As a result for any $N$, we have the asymptotic
convergence in trace distance%
\begin{equation}
\lim_{\mu}\left\Vert \rho^{n}-\Lambda(\chi_{\mathcal{E}}^{\mu\otimes
n})\right\Vert =0\text{.}%
\end{equation}

We may therefore use the lower semi-continuity of the relative
entropy~\cite{HolevoBOOK}. In fact, we may write
\begin{align}
E_{P}(\rho^{n})  &  =\inf_{\sigma\in\text{PPT}}S(\rho^{n}||\sigma)\nonumber\\
&  \overset{(1)}{\leq}\inf_{\sigma^{\mu}}S\left[  \lim_{\mu}\Lambda
(\chi_{\mathcal{E}}^{\mu\otimes n})~||~\lim_{\mu}\sigma^{\mu}\right]
\nonumber\\
&  \overset{(2)}{\leq}\inf_{\sigma^{\mu}}\underset{\mu\rightarrow+\infty}%
{\lim\inf}~S\left[  \Lambda(\chi_{\mathcal{E}}^{\mu\otimes n})~||~\sigma^{\mu
}\right] \nonumber\\
&  \overset{(3)}{\leq}\inf_{\sigma^{\mu}}\underset{\mu\rightarrow+\infty}%
{\lim\inf}~S\left[  \Lambda(\chi_{\mathcal{E}}^{\mu\otimes n})~||~\Lambda
(\sigma^{\mu})\right] \nonumber\\
&  \overset{(4)}{\leq}\inf_{\sigma^{\mu}}\underset{\mu\rightarrow+\infty}%
{\lim\inf}~S\left(  \chi_{\mathcal{E}}^{\mu\otimes n}~||~\sigma^{\mu}\right)
\nonumber\\
&  \overset{(5)}{=}E_{P}(\chi_{\mathcal{E}}^{\otimes n}), \label{Tocomb2}%
\end{align}
where: (1)$~\sigma^{\mu}$ is a sequence of PPT states such that $\Vert
\sigma-\sigma^{\mu}\Vert\overset{\mu}{\rightarrow}0$ for some PPT $\sigma$;
(2)~we use the lower semi-continuity of the relative entropy~\cite{HolevoBOOK}%
; (3)~we use that $\Lambda(\sigma^{\mu})$ are specific types of converging PPT
sequences; (4)~we use the monotonicity of the relative entropy under
trace-preserving LOCCs; and (5)~we use the definition of RPPT for asymptotic
states of Eq.~(\ref{asyBBB}).

Combining Eqs.~(\ref{Tocomb1}) and~(\ref{Tocomb2}), we then derive%
\begin{equation}
nR_{n,d,N}\leq E_{P}(\chi_{\mathcal{E}}^{\otimes n})+f(\varepsilon
,d^{nR_{n,d,N}}),
\end{equation}
for any $n$, $d$ and $N$. We can compute the extension of Eq.~(\ref{Toextend}%
), which is
\begin{equation}
\lim_{n\rightarrow\infty}R_{n,d,N}\leq\frac{E_{P}^{\infty}\left(
\chi_{\mathcal{E}}\right)  }{1-\frac{\varepsilon}{2}\mathrm{log}_{2}d}.
\end{equation}
For $\varepsilon\rightarrow0$ (weak converse), we obtain $\lim_{n\rightarrow
\infty}R_{n,d,N}\leq E_{P}^{\infty}\left(  \chi_{\mathcal{E}}\right)  $ and
the optimization over the original protocols $\mathcal{P}$ automatically leads
to the upper bound%
\begin{equation}
Q_{2}(\mathcal{E}|d,N):=\sup_{\mathcal{P}}\lim_{n\rightarrow\infty}%
R_{n,d,N}\leq E_{P}^{\infty}\left(  \chi_{\mathcal{E}}\right)  ~.
\end{equation}

Since the right hand side does not depend on the input energy constraint $N$
and the output truncated dimension $d$, we may extend it to the supremum,
i.e.,
\begin{equation}
Q_{2}(\mathcal{E})=\sup_{d,N}Q_{2}(\mathcal{E}|d,N)\leq E_{P}^{\infty}\left(
\chi_{\mathcal{E}}\right)  .
\end{equation}


\begin{thebibliography}{99}
\bibitem {NoteLOCC}In general, a LOCC simulation of a channel is a
representation of the channel $\mathcal{E}$ where it is replaced by an LOCC
(or a sequence of LOCCs) applied to the input state $\rho$ and a suitable
resource state $\sigma$ (which may be finite or infinite
dimensional)~\cite{Stretching}. A preliminary version of the LOCC simulation
was proposed in Ref.~\cite{HoroTEL}. However, because the latter does not
involve sequences of LOCCs and only considers finite dimensional $\sigma$, it
cannot simulate the amplitude damping channel, i.e., it cannot cover all the
DV channels.
\bibitem {ProvaCV}The proof for continuous-variable channels is more
complicated and is not reported here. It is based on a suitable extension of
the squashed entanglement functional to account for asymptotic states.
\bibitem {remark}For technical reasons connected with the proof of the
single-path bound, we cannot consider multi-copy REE quantities or the
regularisation $E_{R}^{\infty}$ in Eq.~(\ref{cutset}).
\end{thebibliography}
\end{document}